\documentstyle[12pt,epsf]{article}
\makeatletter
\def\@addtopreamble#1{\xdef\@preamble{\@preamble #1}}
\def\@testpach{\@chclass
 \ifnum \@lastchclass=6 \@ne \@chnum \@ne \else
  \ifnum \@lastchclass=7 5 \else
   \ifnum \@lastchclass=8 \tw@ \else
    \ifnum \@lastchclass=9 \thr@@
   \else \z@
   \ifnum \@lastchclass = 10 \else
   \edef\@nextchar{\expandafter\string\@nextchar}%
   \@chnum
   \if \@nextchar c\z@ \else
    \if \@nextchar l\@ne \else
     \if \@nextchar r\tw@ \else
   \z@ \@chclass
   \if\@nextchar |\@ne \else
    \if \@nextchar !6 \else
     \if \@nextchar @7 \else
      \if \@nextchar <8 \else
       \if \@nextchar >9 \else
  10
  \@chnum
  \if \@nextchar m\thr@@\else
   \if \@nextchar p4 \else
    \if \@nextchar b5 \else
   \z@ \@chclass \z@ \@preamerr \z@ \fi \fi \fi \fi
   \fi \fi  \fi  \fi  \fi  \fi  \fi \fi \fi \fi \fi \fi}
\def\@xexpast#1*#2#3#4\@@{%
   \@tempcnta #2
   \toks@={#1}\@temptokena={#3}%
   \let\the@toksz\relax \let\the@toks\relax
   \def\@tempa{\the@toksz}%
   \ifnum\@tempcnta >0 \@whilenum\@tempcnta >0\do
     {\edef\@tempa{\@tempa\the@toks}\advance \@tempcnta \m@ne}%
       \let \@tempb \@xexpast \else
       \let \@tempb \@xexnoop \fi
   \def\the@toksz{\the\toks@}\def\the@toks{\the\@temptokena}%
   \edef\@tempa{\@tempa}%
   \expandafter \@tempb \@tempa #4\@@}
\def\prepnext@tok{\advance \count@ \@ne
   \toks\count@{}}
\def\save@decl{\toks\count@ \expandafter{\@nextchar}}
\def\insert@column{%
   \the@toks \the \@tempcnta
   \ignorespaces \@sharp \unskip
   \the@toks \the \count@ \relax}
\newdimen\col@sep
\def\@acol{\@addtopreamble{\hskip\col@sep}}
\def\@mkpream#1{\gdef\@preamble{}\@lastchclass 4 \@firstamptrue
   \let\@sharp\relax \let\@startpbox\relax \let\@endpbox\relax
   \@xexpast #1*0x\@@
   \count@\m@ne
   \let\the@toks\relax
   \prepnext@tok
   \expandafter \@tfor \expandafter \@nextchar
    \expandafter :\expandafter =\@tempa \do
   {\@testpach
   \ifcase \@chclass \@classz \or \@classi \or \@classii
     \or \save@decl \or \or \@classv \or \@classvi
     \or \@classvii \or \@classviii  \or \@classix
     \or \@classx \fi
   \@lastchclass\@chclass}%
   \ifcase\@lastchclass
   \@acol \or
   \or
   \@acol \or
   \@preamerr \thr@@ \or
   \@preamerr \tw@ \@addtopreamble\@sharp \or
   \or
   \else  \@preamerr \@ne \fi
   \def\the@toks{\the\toks}}
\def\@classx{%
  \ifcase \@lastchclass
  \@acolampacol \or
  \@addamp \@acol \or
  \@acolampacol \or
  \or
  \@acol \@firstampfalse \or
  \@addamp
  \fi}
\def\@classz{\@classx
   \@tempcnta \count@
   \prepnext@tok
   \@addtopreamble{\ifcase \@chnum
      \hfil
      \d@llarbegin
      \insert@column
      \d@llarend \hfil \or
      \d@llarbegin \insert@column \d@llarend \hfil \or
      \hfil\kern\z@ \d@llarbegin \insert@column \d@llarend \or
   $\vcenter
   \@startpbox{\@nextchar}\insert@column \@endpbox $\or
   \vtop \@startpbox{\@nextchar}\insert@column \@endpbox \or
   \vbox \@startpbox{\@nextchar}\insert@column \@endpbox
  \fi}\prepnext@tok}
\def\@classix{\ifnum \@lastchclass = \thr@@
       \@preamerr \thr@@ \fi
       \@classx}
\def\@classviii{\ifnum \@lastchclass >\z@
      \@preamerr 4\@chclass 6 \@classvi \fi}
\def\@arrayrule{\@addtopreamble \vline}
\def\@classvii{\ifnum \@lastchclass = \thr@@
   \@preamerr \thr@@ \fi}
\def\@classvi{\ifcase \@lastchclass
      \@acol \or
      \@addtopreamble{\hskip \doublerulesep}\or
      \@acol \or
      \@classvii
      \fi}
\def\@classii{\advance \count@ \m@ne
   \save@decl\prepnext@tok}
\def\@classv{\save@decl
   \@addtopreamble{\d@llarbegin\the@toks\the\count@\relax\d@llarend}%
   \prepnext@tok}
\def\@classi{\@classvi
   \ifcase \@chnum \@arrayrule \or
      \@classv \fi}
\def\@startpbox#1{\bgroup
  \hsize #1 \@arrayparboxrestore
   \everypar{%
      \vrule \@height \ht\@arstrutbox \@width \z@
      \everypar{}}%
   }
\def\@endpbox{\@finalstrut\@arstrutbox \egroup}
\def\@array[#1]#2{%
  \@tempdima \ht \strutbox
  \advance \@tempdima by\extrarowheight
  \setbox \@arstrutbox \hbox{\vrule
             \@height \arraystretch \@tempdima
             \@depth \arraystretch \dp \strutbox
             \@width \z@}%
  \begingroup
  \@mkpream{#2}%
  \xdef\@preamble{\ialign \@halignto
                  \bgroup \@arstrut \@preamble
                          \tabskip \z@ \cr}%
  \endgroup
  \@arrayleft
  \if #1t\vtop \else \if#1b\vbox \else \vcenter \fi \fi
  \bgroup
  \let \@sharp ##\let \protect \relax
  \lineskip \z@
  \baselineskip \z@
  \m@th
  \let\\\@arraycr \let\tabularnewline\\\let\par\@empty \@preamble}
\newdimen \extrarowheight
\def\@arstrut{\unhcopy\@arstrutbox}
\def\@arraycr{{\ifnum 0=`}\fi
  \@ifstar \@xarraycr \@xarraycr}
\def\@xarraycr{\@ifnextchar [%
  \@argarraycr {\ifnum 0=`{\fi}\cr}}
\def\@argarraycr[#1]{\ifnum0=`{\fi}\ifdim #1>\z@
  \@xargarraycr{#1}\else \@yargarraycr{#1}\fi}
\def\@xargarraycr#1{\unskip
  \@tempdima #1\advance\@tempdima \dp\@arstrutbox
  \vrule \@depth\@tempdima \@width\z@ \cr}
\def\@yargarraycr#1{\cr\noalign{\vskip #1}}
\def\multicolumn#1#2#3{%
   \multispan{#1}\begingroup
   \def\@addamp{\if@firstamp \@firstampfalse \else
                \@preamerr 5\fi}%
   \@mkpream{#2}\@addtopreamble\@empty
   \endgroup
   \def\@sharp{#3}%
   \@arstrut \@preamble
   \null
   \ignorespaces}
\let\d@llarbegin\begingroup
\let\d@llarend\endgroup
\def\array{\col@sep\arraycolsep
  \def\d@llarbegin{$}\let\d@llarend\d@llarbegin\gdef\@halignto{}%
  \@tabarray}
\def\@tabarray{\@ifnextchar[{\@array}{\@array[c]}}
\def\tabular{\gdef\@halignto{}\@tabular}
\expandafter\def\csname tabular*\endcsname#1{%
      \gdef\@halignto{to#1}\@tabular}
\def\@tabular{%
  \leavevmode
  \hbox \bgroup $\col@sep\tabcolsep \let\d@llarbegin\begingroup
                                    \let\d@llarend\endgroup
  \@tabarray}
\def\endarray{\crcr \egroup \egroup \gdef\@preamble{}}
\def\endtabular{\endarray $\egroup}
\expandafter\let\csname endtabular*\endcsname=\endtabular
\let\@ampacol=\relax        \let\@expast=\relax
\let\@arrayclassiv=\relax   \let\@arrayclassz=\relax
\let\@tabclassiv=\relax     \let\@tabclassz=\relax
\let\@arrayacol=\relax      \let\@tabacol=\relax
\let\@tabularcr=\relax      \let\@@endpbox=\relax
\let\@argtabularcr=\relax   \let\@xtabularcr=\relax
\def\@preamerr#1{\def\@tempd{{..} at wrong position: }%
   \PackageError{array}{%
   \ifcase #1 Illegal pream-token (\@nextchar): `c' used\or %
    Missing arg: token ignored\or                           %
    Empty preamble: `l' used\or                             %
    >\@tempd token ignored\or                               %
    <\@tempd changed to !{..}\or                            %
    Only one column-spec. allowed.\fi}\@ehc}                %
\def\newcolumntype#1{%
  \edef\NC@char{\string#1}%
  \@ifundefined{NC@find@\NC@char}%
    {\@tfor\next:=<>clrmbp@!|\do{\if\noexpand\next\NC@char
        \PackageWarning{array}%
                       {Redefining primitive column \NC@char}\fi}%
     \NC@list\expandafter{\the\NC@list\NC@do#1}}%
    {\PackageWarning{array}{Column \NC@char\space is already defined}}%
  \@namedef{NC@find@\NC@char}##1#1{\NC@{##1}}%
  \@ifnextchar[{\newcol@{\NC@char}}{\newcol@{\NC@char}[0]}}
\def\newcol@#1[#2]#3{\expandafter\@reargdef
     \csname NC@rewrite@#1\endcsname[#2]{\NC@find#3}}
\def\NC@#1{%
  \@temptokena\expandafter{\the\@temptokena#1}\futurelet\next\NC@ifend}
\def\NC@ifend{%
  \ifx\next\relax
    \else\@tempswatrue\expandafter\NC@rewrite\fi}
\def\NC@do#1{%
  \expandafter\let\expandafter\NC@rewrite
    \csname NC@rewrite@\string#1\endcsname
  \expandafter\let\expandafter\NC@find
    \csname NC@find@\string#1\endcsname
  \expandafter\@temptokena\expandafter{\expandafter}%
        \expandafter\NC@find\the\@temptokena#1\relax}
\def\showcols{{\def\NC@do##1{\let\NC@do\NC@show}\the\NC@list}}
\def\NC@show#1{%
  \typeout{Column #1\expandafter\expandafter\expandafter\NC@strip
  \expandafter\meaning\csname NC@rewrite@#1\endcsname\@@}}
\def\NC@strip#1:#2->#3 #4\@@{#2 -> #4}
\newtoks\NC@list
\newcolumntype{*}[2]{}
\long\@namedef{NC@rewrite@*}#1#2{%
  \count@#1
  \loop
  \ifnum\count@>\z@
  \advance\count@\m@ne
  \@temptokena\expandafter{\the\@temptokena#2}%
  \repeat
  \NC@find}
\let\@xexpast\relax
\let\@xexnoop\relax
\def\save@decl{\toks \count@ = \expandafter\expandafter\expandafter
                  {\expandafter\@nextchar\the\toks\count@}}
\def\@mkpream#1{\gdef\@preamble{}\@lastchclass 4 \@firstamptrue
   \let\@sharp\relax \let\@startpbox\relax \let\@endpbox\relax
   \@temptokena{#1}\@tempswatrue
   \@whilesw\if@tempswa\fi{\@tempswafalse\the\NC@list}%
   \count@\m@ne
   \let\the@toks\relax
   \prepnext@tok
   \expandafter \@tfor \expandafter \@nextchar
    \expandafter :\expandafter =\the\@temptokena \do
   {\@testpach
   \ifcase \@chclass \@classz \or \@classi \or \@classii
     \or \save@decl \or \or \@classv \or \@classvi
     \or \@classvii \or \@classviii
     \or \@classx
     \or \@classx \fi
   \@lastchclass\@chclass}%
   \ifcase\@lastchclass
   \@acol \or
   \or
   \@acol \or
   \@preamerr \thr@@ \or
   \@preamerr \tw@ \@addtopreamble\@sharp \or
   \or
   \else  \@preamerr \@ne \fi
   \def\the@toks{\the\toks}}
\let\@classix\relax
\def\@classviii{\ifnum \@lastchclass >\z@\ifnum\@lastchclass=\tw@\else
      \@preamerr 4\@chclass 6 \@classvi \fi\fi}
\def\@classv{\save@decl
   \expandafter\NC@ecs\@nextchar\extracolsep{}\extracolsep\@@@
   \@addtopreamble{\d@llarbegin\the@toks\the\count@\relax\d@llarend}%
   \prepnext@tok}
\def\NC@ecs#1\extracolsep#2#3\extracolsep#4\@@@{\def\@tempa{#2}%
  \ifx\@tempa\@empty\else\toks\count@={#1\tabskip#2\relax#3}\fi}
\def\@tabarray{\@ifnextchar[{\@@array}{\@@array[c]}}
\let\@@array\@array
\def\endarray{\crcr \egroup \egroup \@arrayright \gdef\@preamble{}}
\let\@arrayleft\@empty
\let\@arrayright\@empty
\newlength{\extratabsurround}
\newlength{\backup@length}
\newcommand{\firsthline}{%
  \multicolumn1c{%
    \global\backup@length\ht\@arstrutbox
    \global\advance\backup@length\dp\@arstrutbox
    \advance\backup@length\extratabsurround
    \rule\z@\backup@length
    }\\[-\backup@length]\hline
}
\newcommand{\lasthline}{\hline\multicolumn1c{}%
    \global\backup@length-\ht\@arstrutbox
    \global\advance\backup@length\extratabsurround
    \\[\backup@length]%
}
\makeatother
\renewcommand{\thefootnote}{\fnsymbol{footnote}}
\begin{document}

\hskip 10cm {\sl DESY 95-034}
\vskip.0pt
\hskip 10cm {\sl HD-THEP-94-51}
\vskip.0pt
\hskip 10cm hep-th/9502125
\vskip .7cm
\begin{center}
{\Large \bf Yukawa Couplings for the Spinning Particle
} \\[1ex]
{\Large \bf
and the Worldline Formalism} \\[11ex]
\vskip1cm
 {\large Myriam Mondrag\'on
\footnote{e-mail address Myriam.Mondragon@urz.uni-heidelberg.de},
Lukas Nellen
\footnote{e-mail address Lukas.Nellen@urz.uni-heidelberg.de},
Michael G. Schmidt
\footnote{e-mail address k22@vm.urz.uni-heidelberg.de}
}\\[1.5ex]
  \vskip.1cm
 {\it
  Institut f\"ur Theoretische Physik\\
  Universit\"at Heidelberg\\
  Philosophenweg 16\\
  D-69120 Heidelberg, Germany\medskip\\}
\vskip.2cm
 {\large Christian Schubert
   \footnote{e-mail address schubert@qft2.physik.hu-berlin.de}}\\[1.5ex]
 {\it
  Humboldt Universit\"at zu Berlin\\
  Institut f\"ur Physik\\
  Invalidenstr. 110\\
   D-10115 Berlin, Germany\medskip \\

\vskip .7cm
  {\bf February 1995}
   }\\[14ex]
 {\large \bf Abstract}
\end{center}
\begin{quotation}
We construct the world-line action for a Dirac
particle coupled to a classical scalar or pseudo-scalar
background field. This action can be used to compute
loop diagrams and the effective action in the Yukawa
model using the world-line path-integral formalism for
spinning particles.
\end{quotation}
\clearpage
\renewcommand{\thefootnote}{\protect\arabic{footnote}}
\pagestyle{plain}

\newcommand{\rmd}{{\rm d}}

\setcounter{footnote}{0}

In the framework of the Bern-Kosower
formalism~\cite{bk:prl66,bk:npb379,bd:npb379}, derived from
string theory, it is
possible to
reproduce results of tree and
one-loop field theory calculations in a compact
and elegant form which has its origin in the simplicity of the string
theory perturbation expansion.
Many of
the results for one-loop calculations can also be understood through
world-line path integrals~\cite{str:npb385,ss:plb318}.
This approach to quantum field theory can be generalised to
multi-loop calculations~\cite{ss:plb331,ss:hd-thep-94-25}.
For earlier applications of world-line path integrals to quantum field
theory see also~\cite{fey:pr80,fey:pr84,hjs:prd16,alv:cmp90}.

In the following we construct the action for the Yukawa coupling for
the spinning particle in Euclidean space.
While the original Bern-Kosower approach works for Yukawa
couplings~\cite{bddk:npb425}
the correct form of the world-line action was so far
unknown,
even though guesses existed before~\cite{str:npb385}.
Our starting point is the well-known
world-line action for a
massless, spinning particle coupled to a gauge
field~\cite{bdh:npb118,bm:ap104}.
By dimensional reduction of the gauge coupling we obtain the form of
the Yukawa coupling.
For the case of a constant background field we recover the action for the
massive, spinning particle as a special case.
It turns out that we can generalise this to include pseudo-scalar
couplings as well as scalar couplings.

We present some applications to the calculation of one-loop amplitude
both to reproduce the result of Feynman diagram calculations and to
get a variant of the heat-kernel expansion for the one-loop effective
action.

Our starting point is the first-quantized description of a Dirac
particle~\cite{bdzdh:pl64b,bdh:npb118,bm:ap104}, given by a
supersymmetric action in one dimension.
Such an action can be formulated in
a compact way using superfield notation. Furthermore, this
notation ensures the supersymmetry of the action, an important point
since we want to add new couplings without loosing supersymmetry.

Our interest lies in the application of the world-line formalism to
loop calculations with external scalar fields.
For this
reason we do
not have to worry about the problem of boundary conditions for the
fermions like in the derivation of the Dirac
propagator from the world-line action~\cite{ht:ap143,fm:npb306}.

In the superfield formulation~\cite{bdzdh:pl64b}, the world-line
parameter $\tau{}$ is
supplemented by an anti-commuting Grassmann parameter $\theta{}$ to
form a two-dimensional superspace. In this formulation, the free
spinning particle is described using world-line superfields
with a space-time vector index
\begin{equation}
  X^\mu{}(\tau{},\theta{})
    = x^\mu{}(\tau{}) + \theta{}\sqrt{e}\,\psi{}^\mu{}(\tau{}),
\label{superfield}
\end{equation}
where $x$ is a normal commuting number, and $\theta{}$ is a Grassmann
variable.

To be able to write down a reparametrisation invariant action we need
to introduce the super-einbein~\cite{bdzdh:pl64b}
\begin{equation}
  \Lambda{} = e + \theta{}\sqrt{e}\,\chi{}.
\end{equation}
In curved superspace, two independent derivatives exist:
\begin{eqnarray}
  D_\theta{} &=& \Lambda{}^{-1/2}
                 \left(\frac{\partial}{\partial\theta{}}
                       - \theta{}\frac{\partial}{\partial\tau{}}\right),
                       \nonumber\\
  D_\tau{} &=& \Lambda{}^{-1}\frac{\partial}{\partial\tau{}}.
\end{eqnarray}
Using these ingredients, the world-line action for a free, spinning
particle is
\begin{equation}
  S_0 = \frac{1}{2} \int \rmd\tau{}\rmd\theta{}\,\Lambda{}^{1/2}
                       D_\tau{}X\cdot{}D_\theta{}X.
\label{free.s}
\end{equation}
In components, this reads
\begin{equation}
  S_0 = \frac{1}{2} \int \rmd\tau{} \left(
                  \frac{\dot x^2}{e}
                  + \frac{1}{e}\chi{}\dot x \psi{}
                  + \psi{}\dot\psi{}
                  \right).
\label{free.c}
\end{equation}

The coupling of the spinning particle to a Yang-Mills field is
well-known~\cite{bdh:npb118}:
\begin{equation}
  S_{\rm YM} = \int \rmd\tau{}\rmd\theta{}\,\Lambda{}^{1/2}
               i g D_\theta X_\mu A^\mu.
\end{equation}
If we take this coupling in five dimensions and analyze it from a
four-dimensional point of view, we find a Dirac-spinor with
both Yang-Mills and Yukawa couplings.
By choosing a background field
such that $A_\mu=0$ for $\mu=0,\dots,3$ we can get a system which has
only one Yukawa coupling.

In the process of dimensional reduction we single out the
fields~$X_5$ and~$A_5$. Furthermore, we notice that $X_5$ only appears
in the
combination~$DX_5$. Therefore we introduce the following
convenient definitions:
\begin{equation}
  \bar X \equiv \Lambda^{1/2} D_\theta X_5,\qquad \Phi \equiv A^5.
\end{equation}
With these definitions the world-line action for a spinning particle
with a Yukawa coupling is
\begin{equation}
  S_{\rm Y} = S_0 + \frac{1}{2} \int \rmd\tau{}\rmd\theta{}\,\Lambda{}^{1/2}
                        \left(
                        \Lambda^{-1} \bar X D_\theta \bar X
                        + 2 \Lambda^{-1/2} i \lambda \bar X \Phi(X)
                        \right).
\label{yukawa.s}
\end{equation}
To rewrite this expression in components we expand the superfield $\bar X$,
which is
fermionic, and the scalar field $\Phi$ as
\begin{equation}
  \bar X = \sqrt{e}\,\psi_5 + \theta x_5
  \qquad\mbox{and}\qquad
  \Phi(X) = \Phi(x) + \theta \sqrt{e}\, \psi^\mu \partial_\mu \Phi(x).
\end{equation}
(This expansion defines~$x_5$ for the rest of this letter, not
eq.~(\ref{superfield}).
We apologise for the confusion.)
In component notation, the action~(\ref{yukawa.s}) is
\begin{equation}
  \begin{array}[b]{r}
  \displaystyle
  S_{\rm Y} = \frac{1}{2}\int\rmd\tau\left\{
                          \frac{\dot x^2}{e} + \frac{x_5^2}{e}
                          + \psi\cdot\dot\psi
                          + \psi_5\dot\psi_5
                          + \chi\left(
                              \frac{1}{e}\dot x \psi - x_5\psi_5
                            \right)
                          \right.\qquad\\
  \displaystyle \left.\vphantom{\frac{\dot x^2}{e}}
                          {} + 2 i \lambda \Big(
                              x_5 \Phi(x)
                              - e \psi_5 \psi \cdot \partial \Phi(x)
                            \Big)
                          \right\}.

  \end{array}
  \label{yukawa.c}
\end{equation}
To introduce a mass term for the fermions, all we have to do is to
add a constant piece to the scalar field.
If we shift the scalar field $\Phi$ by a
constant $\Phi\to \Phi+m/\lambda$ we introduce a mass term into our
action. This procedure introduces a term of the form~$2ix_5m$ which
turns out to be an inconvenience when we want to construct the
perturbation expansion (eq.~(\ref{eff-act})).
To eliminate this term, we shift~$x_5\to x_5-iem$ and get
\begin{equation}
  \begin{array}[b]{@{}>{\displaystyle}r@{}}
    S_{\rm Y} = \frac{1}{2}\int\rmd\tau\left\{
                                       \frac{\dot x^2}{e} + \frac{x_5^2}{e}
                          + e m^2
                          + \psi\cdot\dot\psi
                          + \psi_5\dot\psi_5
                          + \chi\left(
                              \frac{1}{e}\dot x \psi - (x_5-iem)\psi_5
                            \right)
                          \right.\quad\\
  \displaystyle \left.\vphantom{\frac{\dot x^2}{e}}
                          {} + 2 \lambda \Big(
                              (i x_5 + em) \Phi(x)
                              - i e \psi_5 \psi \cdot \partial \Phi(x)
                            \Big)
                          \right\}.
  \end{array}
  \label{yukawa.m}
\end{equation}
{}From this we can integrate out the auxiliary field $x_5$,{\it i.~e.,}
we eliminate it using its equation of motion:%
\footnote{This action differs from the action proposed
in~\cite{str:npb385}.
Starting from this suggestion,
we were not able to perform
the calculations analogous to our one-loop examples presented here.
}
\begin{equation}
  \begin{array}[b]{r@{}l}
    \displaystyle
    S_{\rm Y} = \frac{1}{2}\int\rmd\tau\left\{
  \vphantom{\frac{\dot x^2}{e}}\right.&\displaystyle
                    \frac{\dot x^2}{e}
                    + \psi\cdot\dot\psi
                    + \psi_5\dot\psi_5
                    + e m^2 + 2 e \lambda m \Phi + e \lambda^2 \Phi^2
  \\ &\displaystyle \left. \vphantom{\frac{\dot x^2}{e}}
                    - 2 i \lambda e \psi_5 \psi \cdot \partial \Phi(x)
                    + \chi \left(
                        \frac{1}{e}\dot x \psi
                        + i e m \psi_5 + i e \lambda \psi_5 \Phi
                      \right)
                   \right\}.
  \end{array}
  \label{mass}
\end{equation}
If we set $\Phi=0$, we recover the action for the {\em massive\/}, spinning
particle in component notation~\cite{bdzdh:pl64b,bdh:npb118} or,
from eq.~(\ref{yukawa.s}), the action in superfield
language~\cite{hht:cqg5}.

So far, we discussed the scalar coupling
$i\lambda\Phi\bar\psi\psi$. It is also possible to find a world-line
action which reproduces the one-loop effective action for the Yukawa
model with the pseudo-scalar coupling
$\lambda'\Phi'\bar\psi\gamma_5\psi$. The only change necessary is the
introduction of another fermionic superfield $X'=x_6+\theta\sqrt{e}\,\psi_6$
for the new
interaction term. The presence of the two fields~$\bar X$ and~$X'$ ensures
that the terms generated from the scalar and pseudo-scalar
interactions do not mix --- something that the~$\gamma_5$-matrix
ensures in the standard
field theory treatment.

For a massive, scalar field with a pseudo-scalar coupling the resulting
action is then
\begin{equation}
  \begin{array}[b]{r}
\displaystyle
  S_{\rm PS} = \frac{1}{2}\int\rmd\tau\rmd\theta\,\Lambda^{1/2} \left(
      D_\tau X\cdot D_\theta X
      + \Lambda^{-1} \bar X D_\theta \bar X
      + \Lambda^{-1} X' D_\theta X'
\right.\qquad\\
\displaystyle\left.
      {} + \Lambda^{1/2} im\bar X
      + \Lambda^{1/2} i\lambda'X'\Phi'(X)
    \right).
  \end{array}
  \label{ps.s}
\end{equation}
In components (in the $\chi=0$ gauge), after the elimination of the
auxiliary fields, this reads
\begin{equation}
    S_{\rm PS} = \frac{1}{2}\int\rmd\tau\left\{
                    \frac{\dot x^2}{e}
                    + \psi\cdot\dot\psi
                    + \psi_6\dot\psi_6
                    + e m^2 + e {\lambda'}^2 {\Phi'}^2
                    - 2 i \lambda' e \psi_6 \psi \cdot \partial \Phi'(x)
                   \right\}.
  \label{ps.c}
\end{equation}
We could also formulate this action analogous to eq.~(\ref{yukawa.m})
or, from eqs.~(\ref{mass}) and~(\ref{ps.c}), write down the action for
a spinning particle which has both scalar and pseudo-scalar Yukawa couplings.

Now that we have a world-line action for a spinning particle with a
Yukawa coupling we want to apply it to some one-loop calculations. The
starting point is always the world-line expression for the one-loop
effective action~\cite{str:npb385,ss:plb318}
\begin{equation}
  \Gamma(\Phi) = -2 \int_0^\infty \frac{\rmd T}{T}
                        \int {\cal D}x {\cal D}x_5
                             {\cal D}\psi {\cal D}\psi_5\,
                             e^{-T S_{\rm Y}}.
  \label{eff-act}
\end{equation}
This can be used both for deriving rules for the calculation of
one-loop $n$-point functions and
as an expression from which one can derive approximations to the
one-loop effective action itself.

For any calculation we need rules how to evaluate the path integral in
eq.~(\ref{eff-act}).
In principle, the path integral also
includes the integration over the fields~$e$ and~$\chi$. Since
infinitesimal changes in those fields are associated with
infinitesimal reparametrisations we have a gauge invariance in our
system. After treating this with standard methods~\cite{dot:npb285} we
are left with the conventional integral over~$T$, where~$T$ labels
inequivalent circles.
This gauge-fixing is the origin of the factor~$\rmd T / T$ in
eq.~(\ref{eff-act}). The free~${\cal D}x$ path-integral is normalised
to~$(4\pi T)^{-d/2}$, the other free path integrals
are~$1$~\cite{dot:npb285,ss:plb331}.
A common~\cite{bdh:npb118,str:npb385,ss:plb318}, convenient
gauge choice is~$e=2$ and~$\chi=0$.

Before proceeding, we separate the centre of mass~$x_0$ from the embedding
coordinate~$x$ as %
\begin{equation}
  x(\tau) \equiv x_0 + y(\tau),
  \qquad\mbox{with}\qquad
  \int_0^T \rmd\tau\, y(\tau) = 0.
\end{equation}
The remaining path integral can be evaluated using standard methods.
{}From the free part we obtain correlation functions for the quantum
fields. And the exponential containing the interaction gets expanded
according to the chosen approximation.
The interaction part is evaluated using Wick contractions with the
correlation functions~\cite{str:npb385,ss:plb318}
\begin{equation}
  \begin{array}[b]{r@{}l@{}r@{}l}
\displaystyle
    \langle y^\mu(\tau_1) y^\nu(\tau_2)\rangle&
\displaystyle
      {} = -g^{\mu\nu} G_B(\tau_1,\tau_2),&\qquad
\displaystyle
      G_B(\tau_1,\tau_2)&
\displaystyle
        {} = \left|\tau_1 - \tau_2\right|
               - \frac{(\tau_1 - \tau_2)^2}{T},\\[10pt]
\displaystyle
    \langle \psi^\mu(\tau_1) \psi^\nu(\tau_2) \rangle&
\displaystyle
      {} = \frac{1}{2} g^{\mu\nu} G_F(\tau_1,\tau_2),&\qquad
\displaystyle
      G_F(\tau_1,\tau_2)&
\displaystyle
        {} = \mathop{\rm sign}(\tau_1 - \tau_2),\\[10pt]
    \multicolumn{4}{c}{\displaystyle
    \langle \psi_5(\tau_1) \psi_5(\tau_2) \rangle%
      = \frac{1}{2} G_F(\tau_1,\tau_2),}\\
  \end{array}
  \label{correlation-fn}
\end{equation}
and
\begin{equation}
    \langle x_5(\tau_1) x_5(\tau_2)\rangle
       = 2\delta(\tau_1,\tau_2).
  \label{correlation-fn5}
  \end{equation}
The Green's functions are the ones on the circle for bosonic fields
with periodic boundary conditions, and for fermions with anti-periodic
boundary conditions. The extra term in the~$G_B$ results from the
background charge required due to the compact nature of the circle.
To find the Green's function~(\ref{correlation-fn5}) for~$x_5$, one
just has to invert the identity operator.

\begin{figure}[tbp]
  \begin{center}
    \leavevmode
    \setlength{\unitlength}{1cm}
    \begin{picture}(7,2)(-3.5,-1)
      \put(0,0){\makebox(0,0){\epsffile{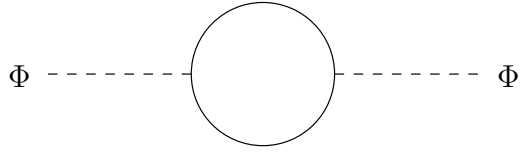}}}
      \put(3.1,0){\makebox(0,0)[l]{$\Phi$}}
      \put(-3.1,0){\makebox(0,0)[r]{$\Phi$}}
    \end{picture}
  \end{center}
  \caption{The two-point function $\Gamma_\Phi^{(2)}$ for the scalar field.}
  \label{fig:2pt}
\end{figure}

As a simple example indicating how to reproduce the results of
a standard Feynman diagram calculation, let us look at the two-point
function for the scalar field in the Yukawa model
(fig.~\ref{fig:2pt}).

{}From the standard expression
\begin{equation}
  (2\pi)^d\delta(p_1+p_2) \Gamma_\Phi^{(2)}(p_1,p_2) =
    \frac{\delta^2}{\delta\Phi(p_1)\delta\Phi(p_2)}
                                     \Gamma(\Phi)\biggr|_{\Phi=0}
  \label{2pt}
\end{equation}
we get the world-line expression for the spinor loop correction to the
scalar two-point function. From the Fourier representation
\begin{equation}
  \Phi(x) = \int \frac{\rmd^d p}{(2\pi)^d} e^{-i p\cdot x} \Phi(p)
  \label{vertex-op}
\end{equation}
the functional differentiation in eq.~(\ref{2pt}) automatically
generates the vertex operator $\Phi \to \exp(-i p\cdot x)$ for the
scalar field. The expression for the scalar propagator correction is
then
\begin{equation}
  \begin{array}[b]{@{}r@{}l@{}}
\displaystyle
    (2\pi)^d\delta(p_1+{}
&\displaystyle
    p_2)
    \Gamma_\Phi^{(2)}(p_1,p_2) =
        -2 \int_0^\infty
        \frac{\rmd T}{T}(4\pi T)^{-d/2}
        e^{-Tm^2} \int \rmd^d x_0\,e^{-i x_0\cdot(p_1+p_2)}\times{}
\\[10pt]&
\displaystyle
        \left[-2\int_0^T\rmd\tau\,\lambda^2
          \langle e^{-ip_1\cdot y(\tau)}
                  e^{-ip_2\cdot y(\tau)}\rangle\right.
\\&
\displaystyle
          \ {}+  \int_0^T\rmd\tau_1\int_0^T\rmd\tau_2
              \left\{ 4\lambda^2m^2
              \langle e^{-ip_1\cdot y(\tau_1)}
                      e^{-ip_2\cdot y(\tau_2)}\rangle\right.
\\&
\displaystyle\left.\left.
          \quad{}- 4 \lambda^2\langle\psi_5(\tau_1)\psi_5(\tau_2)\rangle
                       \langle\psi^\mu(\tau_1)\psi^\nu(\tau_2)\rangle
                       p_{1,\mu} p_{2,\nu}
    \langle e^{-ip_1\cdot y(\tau_1)}
            e^{-ip_2\cdot y(\tau_2)}\rangle
    \right\}\vphantom{\int_0^T}\right].
  \end{array}
  \label{2pt.1}
\end{equation}
Besides the correlation functions from eq.~(\ref{correlation-fn}) we
use the contraction of exponentials
\begin{equation}
  \langle e^{-ip_1\cdot x(\tau_1)} e^{-ip_2\cdot x(\tau_2)}\rangle
    = e^{p_1\cdot p_2 G_B(\tau_1,\tau_2)}
  \label{exponential}
\end{equation}
to evaluate~(\ref{2pt.1}).

The $x_0$-integral produces the momentum conserving $\delta$-function
which we use to set
$p\equiv p_1 = - p_2$.
To evaluate this expression further we use the standard
rescaling%
{}~$\tau_i = Tu_i$%
{}~\cite{str:npb385,ss:plb318,ber:ucla-93-tep-5}.
This way, all the $T$-dependence is displayed explicitly and the
$T$-integration can be done, leading to
\begin{equation}
  \begin{array}[b]{@{}r@{}l@{}}
\displaystyle
    \Gamma_\Phi^{(2)}(p,-p) =
    -2\lambda^2(4\pi)^{\epsilon-2}\left(
\vphantom{\int_0^1}\right.&\displaystyle
        -2\Gamma(\epsilon-1) (m^2)^{1-\epsilon}
\\
&\displaystyle
\left. {}
    + \Gamma(\epsilon) \left[4 m^2 + p^2\right]
      \int_0^1\rmd x \left(m^2 + p^2 x(1-x)\right)^{-\epsilon}
    \right)
  \end{array}
  \label{2pt.st}
\end{equation}
where we introduce $\epsilon = 2-d/2$.

The result of a standard Feynman parameter calculation is
\begin{equation}
  \Gamma_\Phi^{(2)}(p,-p) =
    -2\lambda^2 (4\pi)^{\epsilon-2}\int_0^1 \rmd x\,\left(
      \left[(2+\epsilon) \Gamma(\epsilon-1) + \Gamma(\epsilon)\right]
      (m^2 + p^2 x(1-x))^{1-\epsilon}
    \right)
  \label{2pt.ft}
\end{equation}
from which we can recover result~(\ref{2pt.st}) by judicious
integration by parts.

It is worthwhile to note here that we do not have the straightforward
connection between the $\tau$-intervals of the world-line formalism
and the Feynman~$\alpha$-parameters.
Alternatively, starting from a second-order expression for the fermion
one-loop effective action~\cite{bd:npb379}, it is possible to
reorganise the field theory perturbation series in a manner analogous
to the world-line formalism%
\footnote{We would like to thank D.~C.~Dunbar for discussions on that point.}
(alas, only after performing the
integration over the loop momentum).

The difference in the organisation of the perturbation series will be
even more noticeable in the next example where it is not possible
to isolate the contribution of a single (first-order formalism)
Feynman diagram in the expression generated by the world-line
formalism.

\begin{figure}[tp]
  \begin{center}
    \leavevmode
    \setlength{\unitlength}{1cm}
    \begin{picture}(4,4)(-2,-2)
      \put(0,0){\makebox(0,0){\epsffile{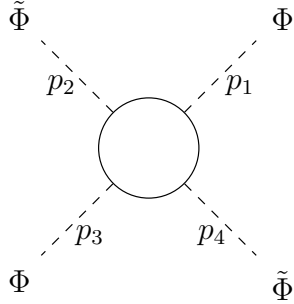}}}
      \put(1.6,1.6){\makebox(0,0)[bl]{$\Phi$}}
      \put(1,1){\makebox(0,0)[tl]{$p_1$}}
      \put(-1.6,-1.6){\makebox(0,0)[tr]{$\Phi$}}
      \put(-1,-1){\makebox(0,0)[tl]{$p_3$}}
      \put(-1.6,1.6){\makebox(0,0)[br]{$\tilde\Phi$}}
      \put(-1,1){\makebox(0,0)[tr]{$p_2$}}
      \put(1.6,-1.6){\makebox(0,0)[tl]{$\tilde\Phi$}}
      \put(1,-1){\makebox(0,0)[tr]{$p_4$}}
    \end{picture}
  \end{center}
  \caption{The mixed four-point function.}
  \label{fig:4pt}
\end{figure}

To illustrate the point we just made, let us look at the four-point
function (fig.~\ref{fig:4pt})
where, in a standard field theory calculation, six Feynman
diagrams contribute.
The world-line formalism generates only one expression which cannot be
divided by restricting the region of integration of the~$\tau$'s
into the contribution corresponding to individual Feynman diagrams.
{}From world-line action~(\ref{ps.c})
we immediately get the expression
\begin{equation}
\newcommand{\VS}{\vphantom{\int_0^T}}
  \begin{array}[b]{>{\displaystyle\qquad\phantom{{}\times\left[\VS\right.}}l}
    \multicolumn{1}{>{\displaystyle\VS}l}{
      \Gamma^{(4)}(p_1,p_2,p_3,p_4)=
        -2\int_0^\infty \frac{\rmd T}{T} (4\pi T)^{-d/2} e^{-Tm^2} \times{}
    }\\
    \multicolumn{1}{>{\displaystyle\VS\qquad}l}{
      \times\left[4{\vphantom{\lambda'}\lambda}^2{\lambda'}^2
        \int_0^T \rmd \tau_1 \int_0^T \rmd \tau_2
        \left\langle
            e^{-ip_1.y_1} e^{-ip_3.y_1} e^{-ip_2.y_2} e^{-ip_4.y_2}
        \right\rangle
        \right.
    }\\
    -8{\vphantom{\lambda'}\lambda}^2{\lambda'}^2
      \int_0^T \rmd \tau_1 \int_0^T \rmd \tau_2 \int_0^T \rmd \tau_3 \times {}
    \\
    \qquad
    \begin{array}[t]{@{}>{\displaystyle\VS}r@{}}
        \left\{(m^2-\langle\psi_{5,1}\psi_{5,2}\rangle
                    \langle\psi_{\mu,1}\psi_{\nu,2}\rangle \, p_1^\mu p_3^\nu)
          \left\langle
              e^{-ip_1.y_1} e^{-ip_3.y_2} e^{-ip_2.y_3} e^{-ip_4.y_3}
          \right\rangle\right.\ %
        \\
        \left.
        {}-\langle\psi_{6,1}\psi_{6,2}\rangle
          \langle\psi_{\mu,1}\psi_{\nu,2}\rangle \, p_2^\mu p_4^\nu
          \left\langle
              e^{-ip_1.y_3} e^{-ip_3.y_3} e^{-ip_2.y_1} e^{-ip_4.y_2}
          \right\rangle
        \right\}
      \end{array}
    \\
    +16{\vphantom{\lambda'}\lambda}^2{\lambda'}^2
      \int_0^T \rmd \tau_1 \int_0^T \rmd \tau_2 \int_0^T \rmd \tau_3
        \int_0^T \rmd \tau_4 \times {}
    \\
    \qquad
    \begin{array}[t]{>{\displaystyle\VS}l}
      \left\{ -m^2 \langle\psi_{6,2}\psi_{6,4}\rangle
          \langle\psi_{\mu,2}\psi_{\nu,4}\rangle \, p_2^\mu p_4^\nu
      \right.
      \\
      \left.\ %
      {} + \langle\psi_{5,1}\psi_{5,3}\rangle
          \langle\psi_{6,2}\psi_{6,4}\rangle
          \langle \psi_{\mu,1} \psi_{\nu,2}
              \psi_{\rho,3} \psi_{\sigma,4} \rangle\,
          p_1^\mu p_2^\nu p_3^\rho p_4^\sigma
      \vphantom{m^2}\right\} \times {}
      \\
      \multicolumn{1}{>{\displaystyle\VS}r@{}}{
        \left.\left\langle
            e^{-ip_1.y_2} e^{-ip_2.y_2} e^{-ip_3.y_3} e^{-ip_4.y_4}
        \right\rangle\vphantom{\int_0^T}\right].
      }
    \end{array}
  \end{array}
  \label{4pt}
\end{equation}
Here we use the short-hand notation $y_i\equiv y(\tau_i)$, and we
omitted the momentum conserving $\delta$-function.
The scalar
fields are attached to the momenta~$p_1$ and~$p_3$ while the
pseudo-scalar fields are on~$p_2$ and~$p_4$.

With the contractions~(\ref{correlation-fn}), it is easy to evaluate
this expression. The $T$-integrations lead again to the
$\Gamma$-functions of dimensional regularisation, and the $G_B$'s
produce polynomials similar to polynomials in Feynman parameters in the
standard field theory calculation. Whereas the $T$-integration is
simple, the remaining $\tau$-integrations are more complicated ---
they are of a form similar to scalar bubble, triangle and box integrals.
To illustrate our point it suffices to
extract the
$1/\epsilon$-term from eq.~(\ref{4pt}). The only divergence sits in
the first term which is%
\begin{equation}
  -\frac{8{\vphantom{\lambda'}\lambda}^2{\lambda'}^2}{(4\pi)^{2-\epsilon}}
    \Gamma(\epsilon)
    \int_0^1 \rmd u\, \left(m^2+(p_1 \cdot p_2
                                 +p_1 \cdot p_4
                                 +p_3 \cdot p_2
                                 +p_3 \cdot p_4) u(1-u) \right)^{-\epsilon}.
  \label{pole}
\end{equation}
This diverges for~$\epsilon\to 0$
as~$-{\vphantom{\lambda'}\lambda}^2{\lambda'}^2 /(2\pi^2\epsilon)$.
The same result is easily obtained from the usual evaluation of
Feynman diagrams in the first-order formalism.
In that case, however,
the contributions of different diagrams enter with different
signs. Such cancelations do not occur in the world-line formalism ---
we immediately get the answer for the sum of several Feynman
diagrams. The point we made before should be  clear after this
example: In general, in the world-line formalism we compute the sum of
a class of Feynman diagrams. It is not always possible to identify the
contribution of a single Feynman diagram by restricting the
integration over the world-line parameters~$\tau_i$.

{}From the effective action~(\ref{eff-act}) we can also generate
very elegantly the
asymptotic heat-kernel expansion of the effective action%
{}~\cite{nep:prd31,ven:npb250,zuk:jmpa18,zuk:prd34}.

Instead of working in momentum space, as we did in the previous
section, we work now in coordinate space. The field $\Phi$ and its
derivative are now expressed through their Taylor expansion
as~\cite{ss:plb318}
\begin{equation}
  \Phi(x(\tau)) = e^{y(\tau)\cdot \partial} \Phi(x_0)
  \label{phi(x)}
\end{equation}
about the centre of mass~$x_0$.
With this we can proceed as before and expand the interaction part of
the exponential
in~(\ref{eff-act}), perform the contractions, and rescale the
integration variables. However, instead of performing the
$T$-integration we expand the exponentials of the form
$\exp(-TG_B(u_i,u_j) \partial_i\cdot\partial_j)$, generated
by~(\ref{exponential}). After arranging the
series by powers of $T$
and performing the $u_i$ integrals
one arrives at a variant~\cite{ss:plb318,fss:zpc64}
of the asymptotic
heat-kernel expansion,
which is particularly well organised~\cite{fhss:hd-thep-94-26}.
In a forthcoming publication~\cite{in-progress} we will discuss in
detail this expansion in the case of the Yukawa model.
The calculation of the expansion of the effective action with
different methods gives us another check that eq.~(\ref{eff-act})
indeed is equivalent to the usual effective action in field theory. We
verified this for the scalar and pseudo-scalar couplings separately
as well as for the mixed case.

We managed to construct the world-line action for the Yukawa model
both for scalar and pseudo-scalar couplings. For some simple examples we
showed how we can reproduce the result of a standard Feynman diagram
calculation using our action in the world-line
formalism.
We performed more calculations to verify the agreement in
other cases which, for the sake of brevity, we do not present in this
letter.

It has been noted in~\cite{ss:plb331} and further exemplified
in~\cite{ss:hd-thep-94-32}
that the world-line path
integral formalism, if  used with a global parametrisation, naturally
offers the possibility of combining diagrams of different topology
into a single master expression.
(For a recent application of this idea, see also~\cite{fl:prd50}).
It will be interesting to see whether this leads to
simplifications in the case of the two-loop box diagrams in
the Yukawa model, which can be treated with the methods
developed in this paper.

\noindent {\bf Acknowledgments:} We would like to acknowledge the
help of D.~Fliegner and discussions with P.~Haberl.

\vspace{1ex}
\noindent
{\bf Note added:} Already in the massive Yukawa model with only a
pseudoscalar coupling,
amplitudes with an odd number of vertices are possible and give rise to
$\epsilon$-tensor terms. We thank D.~Geiser for pointing this out to us.
These terms will be dealt with in a forthcoming publication treating axial
couplings.

\newcommand{\noopsort}[1]{} \newcommand{\switchargs}[2]{#2#1}

\end{document}